\documentclass[aps,prl,article,twocolumn,floatfix, superscriptaddress,longbibliography]{revtex4-2}
\usepackage{lipsum}
\usepackage{graphicx}
\usepackage{epstopdf}
\usepackage{dcolumn}
\usepackage{bm}
\usepackage{physics}
\usepackage{xcolor}
\usepackage{amssymb}
\usepackage{amsmath}
\usepackage{bbold}
\usepackage{soul}
\usepackage[colorlinks=true, linkcolor=blue, urlcolor=blue, citecolor=blue]{hyperref}

\newcommand*{\la}{\langle}
\newcommand*{\ra}{\rangle}

\newcommand{\affFUW}{Faculty of Physics, University of Warsaw, Pasteura 5, 02-093 Warsaw, Poland}
\begin{document}
\preprint{APS/123-QED}

\title{Role of Matter Interactions in Superradiant Phenomena}

\author{Jo\~{a}o Pedro Mendon\c{c}a } 
\email{jpedromend@gmail.com}
\affiliation{\affFUW}
\author{Krzysztof Jachymski}
\email{Krzysztof.Jachymski@fuw.edu.pl}
\affiliation{\affFUW}
\author{Yao Wang}
\email{yao.wang@emory.edu}
\affiliation{Department of Chemistry, Emory University, Atlanta, Georgia 30322, USA}

\date{\today}

\begin{abstract}
   The superradiant phenomenon, usually described by the Dicke model, is a hallmark of strong light-matter interaction. We explore how matter-matter interactions influence this phenomenon by performing ground-state simulations of Dicke-like models with both isotropic and anisotropic spin couplings. We find that Ising-type interactions produce two qualitatively distinct phase boundaries, one of which gives rise to an antiferromagnetic-normal phase connected to the superradiant regime via a first-order phase transition. Under anisotropic couplings, we uncover a strongly correlated phase where in-plane spin order coexists with superradiance, exhibiting sublinear scaling of the photon occupation per site and power-law decay of spin correlations. Furthermore, superradiance can be strengthened by tuning either isotropic or anisotropic interactions, highlighting the role of intrinsic many-body correlations in shaping light-matter quantum phases.
\end{abstract}

\maketitle

Coupling a macroscopic ensemble of quantum emitters to a common electromagnetic mode gives rise to collective phenomena such as photon-mediated entanglement and superradiant photon condensation. These effects underlie emerging technologies in quantum computing, simulation, communication, and sensing\,\cite{Dowling2003,Julsgaard2004,kimble2008quantum,Ritsch2013,Reiserer2015,Findik2021,forn2019ultrastrong}. More recently, these systems have inspired new paradigms such as quantum batteries, which aim to harness collective phenomena like superradiance for fast and efficient energy storage and delivery\,\cite{campaioli2024colloquium,Julia2020}. Substantial progress has been made in this direction, both experimentally\,\cite{Quach2022,joshi2022exp,hu2022optimal,baumann2010dicke} and theoretically\,\cite{Ferraro2018,shi2022qb}. However, capturing the many-body physics of these complex systems remains challenging.

Existing theoretical studies of light-matter hybrid systems commonly rely on two classes of approximations, leading to simplified models such as the Dicke and extended Bose-Hubbard models. The Dicke model retains the photonic degree of freedom but neglects interactions inside the matter system, simplifying it into a collection of non-interacting two-level qubits coupled to a shared photon mode\,\cite{dimer2007proposed,baden2014realization,zhang2018dicke,Ferraro2018, crescente2020ultra, Quach2022}. Although widely adopted in quantum battery and circuit QED studies, the neglected interactions are intrinsic and usually unavoidable in real systems. These interactions originate from multiple mechanisms: in circuit QED, capacitive couplings among qubits give rise to tunable ferromagnetic or antiferromagnetic interactions\,\cite{pashkin2003quantum,Zhang2014,jaako2016ultrastrong,deBernardis2018cavity,viehmann2011superradiant}; in optical cavity QED, dipole-dipole interactions among atoms or molecules lead to short-range spin-exchange terms across optical lattices or tweezer arrays\,\cite{manzoni2017designing,Whitlock2017}; and in quantum batteries, electronic tunneling and molecular interactions generate similar short-range couplings\,\cite{dou2022qb,dou2022xxz,zhang2023qb,wen2025dicke}. These interactions can substantially reshape the emergent phases and alter the performance of quantum devices\,\cite{mivehvar2021cavity}. 

On the other hand, the extended Bose-Hubbard model treats matter degrees of freedom explicitly while integrating out the photonic field, thereby reducing the problem to a competition between short-range and photon-mediated long-range interactions\,\cite{nagy2008self,dogra2016phase,sharma2022quantum,landig2016quantum}. However, this framework cannot capture superradiant effects, which requires retaining the polaritonic wavefunction\,\cite{bezvershenko2021dicke,orso2025self,Tolle2025}. 

\begin{figure}[!t]
    \centering
    \includegraphics[width=0.8\columnwidth]{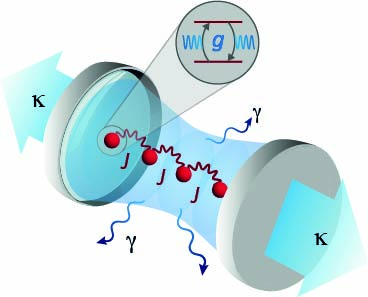}\vspace{-3mm}
    \caption{Schematic of the Dicke-Heisenberg model. The system comprises an ensemble of two-level qubits, embedded in an optical cavity. They interact anisotropically with their nearest neighbors through an exchange interaction $\vb{J}$ and couple collectively to a single cavity mode with strength $g$. Dissipative effects include photon loss at rate $\kappa$ and spontaneous emission at rate $\gamma$, both of which are assumed to be slow compared to the coherent coupling (i.e., $g^2\gg 2\gamma\kappa$).}
    \label{fig:main}
\end{figure}

To overcome the limitations of simplified models and explore the rich many-body physics arising from both light-matter and matter-matter interactions, it is thus needed to incorporate interactions among the matter degrees of freedom [see Fig.~\ref{fig:main}]. In order to simulate the strong-coupling regime efficiently, in this work we develop a hybrid numerical approach that combines a variational polaritonic dressing with an exact many-body solver for the matter subsystem, thereby going beyond mean-field approximations. This framework allows us to treat both light-induced collective effects and interaction-driven correlations on equal footing. Thus, we identify the impact of different types of interactions on the ground-state phases: in the Dicke--Ising model, we observe both first- and second-order phase transitions into the superradiant phase, while in the Dicke-XXZ model, we reveal a coexistence phase featuring XY spin order and superradiance. In both cases, superradiant signatures are significantly enhanced relative to those in the standard Dicke model. 

The extended model we consider, referred to as the Dicke-Heisenberg model, is described by the Hamiltonian
\begin{eqnarray}\label{eq:Hamiltonian}
   \mathcal{H} &= &\omega \, a^\dagger a + \varepsilon \sum_{i=1}^N s_i^z + \frac{2 g}{\sqrt{N}}\sum_{i=1}^N s_i^x(a+a^\dagger)
   \nonumber\\ 
   &&- \sum_{\langle i,j \rangle} \sum_{\alpha=x,y,z} J_\alpha \, s^{(\alpha)}_i \, s^{(\alpha)}_j,
\end{eqnarray}
where the first three terms correspond to the standard Dicke model\,\cite{dicke1954,Safavi2018,Kirton2019} and the final term reflects interactions among the matter degrees of freedom\,\cite{Mattis1963}. Each two-level matter component, separated by energy $\varepsilon$, is encoded as a local spin-$1/2$ operator $s_n^\alpha$. For clarity and generality, we refer to these two-level systems as ``spins'' throughout, although they may represent atomic orbitals, molecular states, or superconducting qubits in experimental implementations. The operators $a$ and $a^\dagger$ denote photon annihilation and creation operators, respectively, with a fixed frequency $\omega$. $g$ is the light-matter interaction strength and $N$ is the system size. We set $\omega=\varepsilon=1$ throughout this paper.

{\it Numerical approach.}
The model in Eq.~\eqref{eq:Hamiltonian} that combines spins and photons poses considerable numerical challenges due to the distinct physical nature of their constituent subsystems. Interacting spins exhibit strong correlations that demand quantum many-body computational techniques \cite{White1992,White1993,Wietek2018,Lauchli2019,Evered2023,Cong2022,Browaeys2020}. In contrast, the photon subsystem inhabits an infinite-dimensional Hilbert space, with occupation numbers that can grow significantly in superradiant or polaronic regimes. Direct truncation of the bosonic space quickly becomes unreliable. However, the photonic field is typically highly coherent, with weak internal entanglement, suggesting that variational methods can effectively capture its wavefunction\,\cite{Guaita2019,Christianen2022}. 

A practical strategy for addressing strong coupling in such composite systems involves a variational unitary transformation, i.e.~the polaritonic dressing, which displaces the photon field in a way that depends on the many-body state of the spins\,\cite{fehske1995hole, Wang2020, wang2021fluctuating}. This transformation rotates the system into a frame where spin and photon wavefunctions are approximately separable. Here, we apply this hybrid variational method to the models defined in Eq.~\eqref{eq:Hamiltonian}. Specifically, the hybrid numerical approach adopted here leverages a non-Gaussian state (NGS) ansatz, $\ket{\psi} = U_\lambda \left(\ket{\psi_\text{ph}}\otimes\ket{\phi}\right)$, where $\ket{\psi_\text{ph}}$ describes the photonic degrees of freedom, $\ket{\phi}$ is a many-body spin wavefunction, and $U_\lambda$ is a non-Gaussian unitary entangling transformation that encodes polaritonic correlations. The photon state $\ket{\psi_\text{ph}}$ is approximated by a Gaussian state $\ket{\psi_{\rm ph}} = e^{i\vb*{R}^T \vb*{\sigma} \vb*{\Delta}_R}e^{-\frac{i}{2}\vb{R}^T \vb*{\xi R}}\ket{0}$, where $\vb{R} = (x,p)^T$ represents the bosonic quadrature vector. The photon displacement is generated by $\vb*{\Delta}_R$ and squeezing is induced by $\vb*{\xi}$\,\cite{Shi2018}. The NGS transformation $U_\lambda$ introduces entanglement between photons and spins:
\begin{equation}
    U_\lambda = \exp{- \lambda \sum_i s_i^{x} (a^\dagger - a)}\,.
\end{equation}
The variational parameter $\lambda$ controls the polaritonic dressing and is determined self-consistently along with the photon variational parameters described below. 

The variational ground state of the Dicke-Heisenberg model is obtained by minimizing the total energy $\mathcal{E}(\vb*{\Delta}_R,\vb*{\xi},\lambda,\ket{\phi})=\bra{\psi}\mathcal{H}\ket{\psi}$. This is accomplished using a self-consistent approach involving two coupled optimization procedures. The first step minimizes the energy with respect to the variational parameters $\vb*{\Delta}_R,\, \vb*{\xi}$, and $\lambda$, which determine the NGS transformation and the photon state. By fixing these variational parameters, the second step numerically calculates the ground state $\ket{\phi}$ of the effective spin Hamiltonian $H_\textrm{eff}(\vb*{\Delta}_R,\vb*{\xi},\lambda) = \bra{\psi_\text{ph}} U_\lambda^\dagger H U_\lambda \ket{\psi_\text{ph}}$, which is renormalized by the above  transformations\,\cite{Takada2003, reja2011phase, Karakuzu2017, Ohgoe2017, reja2018phase, han2024quantum, min2025mixed}, via density matrix renormalization group (DMRG). This numerical simulation is conducted only for the spin subsystem without the necessity of truncating the photonic Hilbert space. Together, these two updates form a single self-consistent iteration, ensuring the energy decrease, which is repeated until convergence. This method ensures that even when the physical photon number becomes large---as is typical in superradiant regimes---the bosonic fluctuations in the transformed frame remain small and numerically tractable. Thus, it enables an accurate and efficient solution for strongly coupled systems with many-body effects (see further analysis in the Supplemental Material\,\cite{SM}).

In this work, we restrict our analysis to a one-dimensional chain with nearest-neighbor spin interactions, a geometry naturally compatible with DMRG due to its low entanglement scaling and efficient tensor network representation\,\cite{itensor}. While exact diagonalization has previously proven efficient for hybrid NGS approaches in electron-phonon systems\,\cite{Wang2020, wang2021phonon}, DMRG offers significant advantages for models such as the Dicke-XXZ, particularly when extrapolating to the thermodynamic limit becomes essential.

\begin{figure}[!b]
    \centering
    \includegraphics[width=0.9\columnwidth]{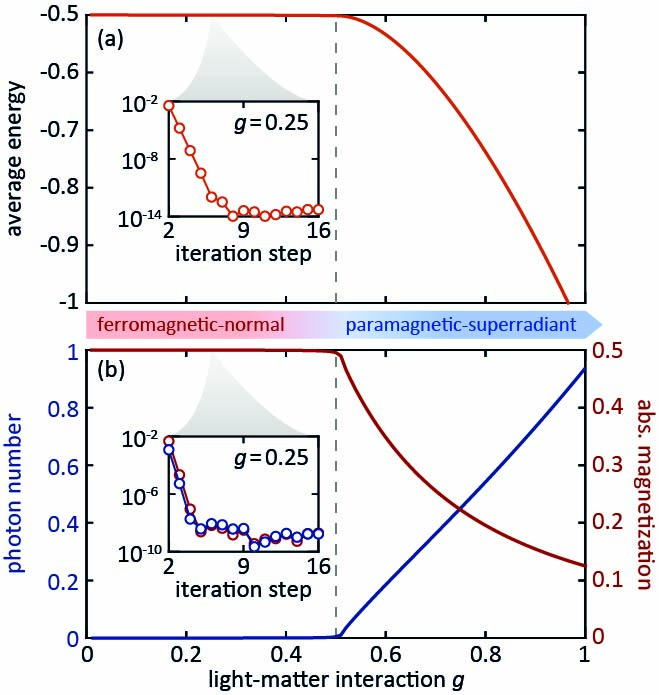}\vspace{-3mm}
    \caption{Ground-state properties of the Dicke model obtained via the hybrid variational method: (a) average energy, (b) average photon occupation and absolute magnetization. Simulations are performed for $N = 200$. The insets show the iteration errors at each step of the self-consistent iterations for the representative case $g = 0.25$ system. The energy is converged to a threshold of $10^{-12}$, while photon number and magnetization are converged to $10^{-8}$.}
    \label{fig:benchmark}
\end{figure}

We first benchmark the hybrid variational framework on the standard Dicke model by setting all spin-spin interactions to zero ($J_\alpha=0$). This serves as a useful test case, allowing direct comparison with analytically known results. We monitor key physical quantities, including the ground-state energy, average photon number, and magnetization. Our simulations capture the well-established superradiant phase transition at $g_c = \sqrt{\omega \varepsilon}/2$ in the large-$N$ limit, as shown in Figs.~\ref{fig:benchmark}(a) and (b).  The magnetization, $M_z = \sum_i \la s^z_i \ra / N$ and per-site photon occupation $\la n \ra/N$ transition sharply across the critical point\,\footnote{As the system maps onto a spin chain, we adopt magnetic terminology to describe observables, although the it may reflect physical quantities in the realization of the matter.}: from a finite $M_z$ and vanishing $\la n \ra/N$ in the normal phase to a vanishing $M_z$ and sizable $\la n\ra/N$ in the superradiant phase. As expected, the total photon number is size independent (equivalently, $\la n \ra/N \approx 0$) in the normal phase, while scaling linearly with $N$ in the superradiant regime, consistent with $\langle n \rangle \propto N$. In addition, because the Dicke model without spin-spin interactions possesses permutation symmetry, its ground state can be solved analytically in the thermodynamic limit via Holstein-Primakoff (HP) transformation\,\cite{Kirton2019,SM}. Our numerical simulations converge toward the analytic coherent-state solution as $N$ increases, validating the accuracy of the hybrid method across both finite-size and thermodynamic regimes.

\begin{figure}[!b]
    \centering
    \includegraphics[width=0.9\columnwidth]{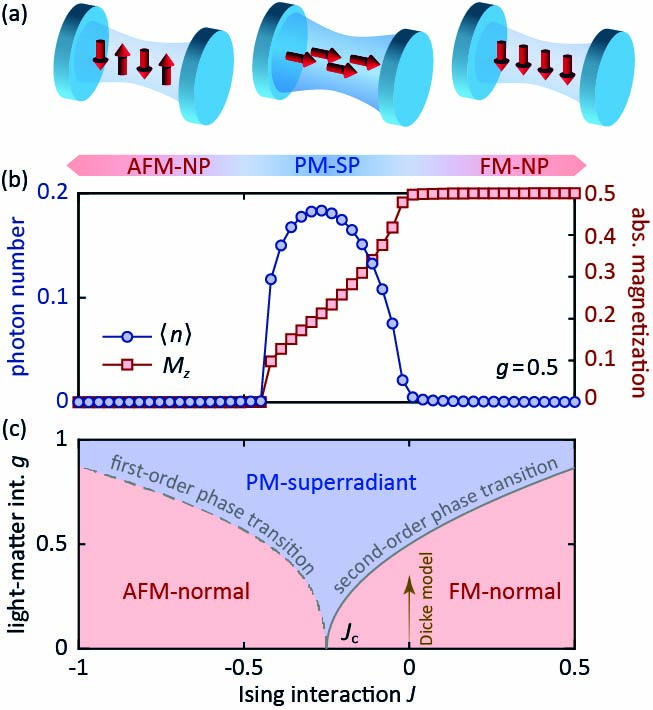}\vspace{-3mm}
    \caption{(a) Representative spin configurations in phases of the Dicke--Ising model. (b) Magnetization and average photon number calculated for $g=0.5$ and $N=100$. (c) Phase diagram of the Dicke--Ising model, where the solid (dashed) lines denote second-order (first-order) transitions between the normal and superradiant phases. The normal phase is further divided by a critical coupling $J_c$ into regions with distinct spin configurations. The vertical arrow marks the Dicke model. }
    \label{fig:DickeIsing_diagram}
\end{figure}

{\it Dicke--Ising model.} Building on the benchmark analysis of the standard Dicke model, we now incorporate spin-spin interactions and investigate the resulting phase diagram of the Dicke--Ising model. These nonperturbative interactions explicitly break permutation symmetry and further reduce residual spin symmetries, which directly affects the superradiant properties and places the model beyond the reach of perturbative or mean-field approximations. To isolate the impact of Ising-like couplings, we set $J_x = J_y = 0$ and $J_z = 4J$. As illustrated in Fig.~\ref{fig:DickeIsing_diagram}(a), the inclusion of spin-spin interactions leads to significant restructuring of the phase diagram, giving rise to three distinct phases described below.

Starting from the standard Dicke limit ($J = 0$), marked by the vertical arrow in Fig.~\ref{fig:DickeIsing_diagram}(c), the critical coupling $g_c$ acquires a $J$-dependence described by $g_c(J) = \sqrt{\omega \varepsilon/4 + \omega J}$\,\cite{SM}. Notably, $g_c$ decreases as $J$ becomes negative and vanishes at $J = -\varepsilon/4$, which defines a critical coupling $J_c$. This separates the normal phase into a ferromagnetic-normal (FM-NP) phase for $J > J_c$ and an antiferromagnetic-normal (AFM-NP) phase for $J < J_c$.

On the FM-NP side ($J > J_c$), the transition into the superradiant phase resembles that of the standard Dicke model. As the light-matter coupling $g$ exceeds the renormalized threshold $g_c(J)$, the system undergoes a continuous phase transition from the normal phase to the superradiant phase, now termed the paramagnetic-superradiant (PM-SP) phase to include the spin configuration. The latter is characterized by a vanishing spin order and finite photon occupation. Alternatively, for a fixed $g$ within the normal phase, decreasing $J$ across the phase boundary leads to a continuous decline in magnetization [see Fig.~\ref{fig:DickeIsing_diagram}(b)], confirming the FM-PM second-order transition  and reflecting the underlying symmetry-breaking mechanism. 
This finding contrasts with a recent conclusion obtained by small-size exact diagonalization with Hilbert space truncation\,\cite{Rohn2020}, which interpreted the transition as first-order. Our simulations extrapolate to the thermodynamic limit and resolve this discrepancy, showing that the FM-NP to PM-SP transition remains second order, consistent with the standard Dicke model. This conclusion is corroborated by a mean-field analysis in the thermodynamic limit, where the impact of the Ising interaction can be effectively absorbed by renormalizing the level splitting as $\varepsilon_{\rm eff} = \varepsilon + 4J$, consistent with the observed $g_c(J)$\,\cite{SM}.

An important consequence of introducing spin-spin interactions is the substantial enhancement of photon number in the superradiant phase. As demonstrated in Fig.~\ref{fig:DickeIsing_diagram}(b), for $g = 0.5$, which coincides with the critical coupling $g_c(J = 0)$ of the standard Dicke model marking the onset of superradiance, the per-site averaged photon number increases markedly with moderate negative values of $J$, reaching nearly $0.2$. This enhancement can be understood as a consequence of the renormalized critical threshold: negative $J$ lowers the effective $g_c$, shifting the system deeper into the superradiant regime at fixed $g$. This mechanism provides a practical route for engineering enhanced superradiant states through moderate spin-spin interactions. However, the effect is not unbounded. As illustrated in Fig.~\ref{fig:DickeIsing_diagram}(b), further decreasing $J$ beyond $-0.27$ causes the photon number to decrease. Thus, an optimal interaction strength exists that maximizes superradiance at a given light-matter coupling. Beyond this point, stronger interactions destabilize the superradiant state due to the onset of a competing phase.

On the AFM-NP side ($J < J_c$), the system exhibits qualitatively different behavior from that of the standard Dicke model. As shown in Fig.~\ref{fig:DickeIsing_diagram}(b), reducing $J$ beyond a secondary threshold ($\sim-0.43$ for $g=0.5$) results in a sharp, discontinuous jump in key observables, indicating a first-order phase transition. Although this phase lacks superradiance, with vanishing photon occupation, it is distinct from both the Dicke normal phase and the FM-NP regime due to its internal spin structure. Specifically, the spin subsystem exhibits long-range antiferromagnetic order, as confirmed by staggered spin correlations $\la (-1)^r s_i^z s_{i+r}^z \ra=1/4$. Because the AFM and superradiant phases break fundamentally different symmetries (translational and parity, respectively), they cannot be adiabatically connected by a continuous phase transition. Therefore, unlike the FM-NP, where the interaction simply renormalizes the level splitting, the AFM-NP to PM-SP reflects a discontinuous change in symmetry of the ground state. This behavior contrasts with a recent variational mean-field study that reported a phase with coexisting AFM order and superradiance\,\cite{Zhang2014}. With careful tuning of parameters and extrapolation to the thermodynamic limit, our simulations do not observe such a coexistence phase. This discrepancy highlights the importance of treating the spin sector exactly: while the photonic sector is often well approximated by a coherent state, strongly correlated spin states may introduce significant effects\,\cite{SM}.

\begin{figure}[!t]
    \centering
    \includegraphics[width=0.9\columnwidth]{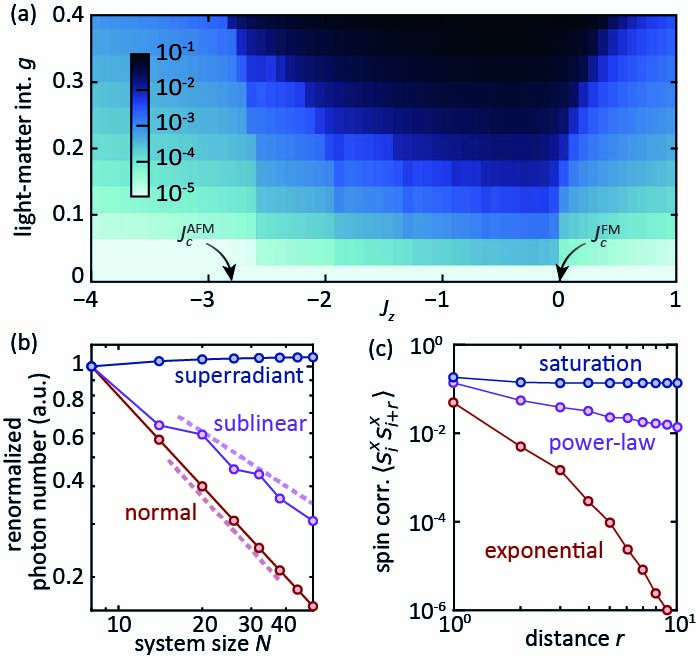}\vspace{-3mm}
    \caption{Phases of the Dicke--XXZ model. (a) Photon number per site $\la n \ra /N$ for various $J_z$ and $g$, obtained use $N=24$. The two arrows indicate the critical couplings at zero light-matter coupling ($g=0$). (b) Scaling laws of $\la n \ra /N$ in different phases: the normal phase (red: $J_z=1$, $g=0.01$) exhibit a $\la n \ra/N \propto 1/N$ decay; the superradiant phase (blue: $J_z=-1.6$, $g=0.4$) shows $\la n \ra \propto N$; a sublinear scaling arises in the intermediate regime (purple: $J_z=-1.6$, $g=0.01$). (c) Spatial profile of the spin-spin correlation $\langle s_i^x s_{i+r}^x \rangle$ for $N=32$. Deep in the normal phase (red: $J_z=-5$, $g=0.01$) it decays exponentially, while the superradiant phase supports long-range order. The intermediate regime displays power-law decay, consistent with quasi-long-range XY-type order [blue and purple the same as in (b)].}
    \label{fig:dickexxz}
\end{figure}

{\it Dicke--XXZ model.}
Beyond isotropic spin-spin couplings, many experimental systems naturally host anisotropic matter-matter interactions, arising from directional dipolar forces and anisotropic tunneling processes\,\cite{micheli2006toolbox,dePaz2013,browaeys2016exp,Whitlock2017}. Motivated by these physical realizations, we further consider the Dicke--XXZ model, where $J_x = J_y = 1$ and $J_z$ is a tunable anisotropy parameter. The distinction between the Dicke--Ising and Dicke--XXZ models primarily inherits from their different ground-state phase diagrams at zero light-matter coupling ($g=0$). Whereas the Dicke-Ising model exhibits a single critical coupling $J_c$ in Fig.~\ref{fig:DickeIsing_diagram}, the XXZ model splits it into two critical couplings $J_c^{\rm FM}=0$ and $J_c^{\rm AFM} \approx -2.8$\,\cite{Justino2012,Niezgoda2020,yang1966I,yang1966II}. The intermediate phase is usually referred to as the gapless XY phase. The introduction of light-matter interaction further broadens up the intermediate phase: a finite photon occupation $\langle n \rangle / N$ emerges beyond these two critical couplings by destabilizing the FM and AFM orders, as shown in Fig.~\ref{fig:dickexxz}(a). 

Within this intermediate region, where in-plane XY correlations dominate and the spin sector is paramagnetic, an infinitesimal light-matter interaction $g$ is sufficient to induce divergence in the total photon number $\langle n \rangle$. While such divergence is a signature of superradiance, its manifestation here differs markedly from the standard Dicke or Dicke--Ising models, where the photon number per site saturates in the superradiant phase [see Figs.~\ref{fig:benchmark} and \ref{fig:DickeIsing_diagram}]. Instead, due to persistent in-plane XY fluctuations, it displays sublinear scaling with system size: $\la n \ra/N \propto N^{-\alpha}$ with $0\leq\alpha <1$ [see Fig.~\ref{fig:dickexxz}(b)]. As the light-matter interaction $g$ increases for a fixed $J_z$, the exponent $\alpha$ decreases continuously toward zero, suggesting a smooth evolution toward the conventional superradiant regime.
For comparison, in the normal FM-NP and AFM-NP phases, we observe $\langle n \rangle / N \propto 1/N$, which implies a constant total photon number but vanishing per-site occupation in the thermodynamic limit.

The coexistence of spin correlations and photon superradiance is another distinctive feature of this intermediate phase. In Fig.~\ref{fig:dickexxz}(c), the in-plane spin correlations $\langle s^x_i s^x_{i+r} \rangle$ decay exponentially in the normal phases, where $z$-axis FM or AFM order dominates. In contrast, within the XY phase, the correlations follow a power-law decay, indicating quasi-long-range order. Although the photon field breaks rotational symmetry in the $xy$-plane by favoring $s^x$ polarization, it does not destroy the XY correlations. Instead, the power-law character persists across the superradiant regime and smoothly evolves into long-range order as $g$ increases, consistent with the Dicke--Ising-like superradiant phase [see Fig.~\ref{fig:dickexxz}(c)].

In conclusion, we have explored how matter-matter interactions shape the phases of strongly coupled light-matter systems described by Dicke-like models, employing a hybrid method that combines variational non-Gaussian states with DMRG. This framework captures both strong light-matter coupling and strong correlations on equal footing, extending the understanding of such systems well beyond mean-field or perturbative treatments (see demonstrations in SM\,\cite{SM}). Our simulations reveal two distinct types of phase boundaries induced by Ising-type interactions: one that adiabatically shifts the conventional superradiant boundary, and another that produces a qualitatively different normal phase characterized by antiferromagnetic order and first-order transition to superradiance. Introducing anisotropic interactions further enriches the phase diagram, leading to a strongly correlated intermediate regime where in-plane spin order coexists with superradiance. These results highlight the intricate effects of matter interactions and the necessity of many-body approaches.

\textit{Acknowledgments.}
J.P.M and K.J. were supported by the Polish National Agency for Academic Exchange (NAWA) via the Polish Returns 2019 program and acknowledge Polish high-performance computing infrastructure PLGrid (HPC Center: ACK Cyfronet AGH) for providing computer facilities and support within computational grant no. PLG/2024/017478. Y.W. acknowledges the support by the U.S. Department of Energy, Office of Science, Basic Energy Sciences, under Early Career Award No.~DE-SC0024524.

\end{document}


\title{Supplemental Material to \\ ``Role of Matter Interactions in Superradiant Phenomena''}

\author{Jo\~{a}o Pedro Mendon\c{c}a } 
\affiliation{\affFUW}
\author{Krzysztof Jachymski}
\affiliation{\affFUW}
\author{Yao Wang}
\affiliation{Department of Chemistry, Emory University, Atlanta, Georgia 30322, USA}

\date{\today}
\maketitle

\section{Analytical Mean-Field Considerations}

In the Dicke–Ising model, the normal phase for $J>J_c$, i.e., the ferromagnetic-normal (FM-NP) phase, can be well captured in the thermodynamic limit using approximate analytical approaches. This phase connects adiabatically to the standard Dicke model. To set the stage, we first revisit the Dicke model, where underlying symmetries predict a second-order phase transition at $g_c=\sqrt{\omega\varepsilon}/2$ in the thermodynamic limit. Although the introduction of Ising interactions formally breaks these symmetries, in the ferromagnetic case the symmetry breaking can be gauged away: the interaction effectively renormalizes the atomic transition frequency $\varepsilon \to \varepsilon_{\rm eff}$, leading only to a shifted phase boundary rather than a qualitatively new phase.

\subsection{Holstein-Primakoff representation}

The absence of direct matter-matter interactions in the standard Dicke model leads to a global spin permutation symmetry: spatial coordinates become irrelevant, and the system reduces to a single collective spin coupled to the uniform single-mode cavity field. The spin Hilbert space is thus spanned by $\{\ket{j,m}, m=-j,-j+1,\cdots, j-1,j\}$, with the collective spin $j=N/2$, reducing the spin degrees of freedom from $2^N$ to $N+1$. Collective spin operators are then defined as $s^\alpha = \sum_j s_j^\alpha$.

This symmetry allows for an analytical treatment via the Holstein-Primakoff (HP) transformation~\cite{holsteinP1940,Kirton2019}, which maps the collective spin onto a bosonic mode:
\begin{equation}\label{eqSM:magnons}
    s^z = - \frac{N}{2} + b^\dagger b, \qquad
    s^x \approx \frac{\sqrt{N}}{2} (b + b^\dagger),
\end{equation}
where the bosonic ladder operators, $b$ and $b^\dagger$ create and annihilate collective spin excitations (magnons). Neglecting terms beyond quadratic order in Eq.~\eqref{eqSM:magnons}, equivalent to taking the $N\gg \langle b^\dagger b\rangle$ limit, yields the quadratic bosonic Hamiltonian:
\begin{equation}
    H^{\rm HP}_{\rm Dicke} = \omega a^\dagger a + \varepsilon b^\dagger b 
    + g (a + a^\dagger)(b + b^\dagger).
\end{equation}

This Hamiltonian represents a system of two coupled harmonic oscillators, i.e., the photon mode and the matter's magnon mode. Diagonalization gives the normal-mode energies
\begin{equation}
    \Omega_{\pm}^2 = \frac{1}{2} \left( \omega^2 + \varepsilon^2 
    \pm \sqrt{(\omega^2 - \varepsilon^2)^2 + 16 g^2 \omega \varepsilon} \right)\,.
\end{equation}
The critical coupling corresponds to softening of the lower-lying mode, i.e., $\Omega_- \to 0$, giving
\begin{equation}\label{eqSM:gcJ}
    g_c = \frac{\sqrt{\omega \varepsilon}}{2}\,,
\end{equation}
which signals a second-order phase transition. This matches the numerical results from our simulations and connects directly to the mean-field theory (MFT), where the Dicke ground state is represented across both normal and superradiant phases. 

Within a variational framework, the MFT is recovered by minimizing the variational energy under a (separable) coherent ansatz:
\begin{equation}\label{eqSM:pure_variational}
    \ket{\psi} = e^{-i\Delta_x p} \ket{0} \otimes e^{-i\phi s^y} \ket{j,-j}\,,
\end{equation}
with the photonic part being described by 
\begin{equation}\label{eqSM:DxMFT}
    \Delta_x = -\frac{ g'\langle s^x \rangle }{\omega},
\end{equation}
where $g'=2g/\sqrt{j}$, and the spin configurations being described by the magnon excitation amplitude
\begin{align}
    \phi =
    \begin{cases}
        0 &\qquad{g<g_c} ,\\
        \cos^{-1}({g_c}^2/g^2) &\qquad{g>g_c}\,.
    \end{cases}
\end{align}
In the strong-coupling limit $g \gg \omega,\varepsilon$, we obtain $\phi=\pi/2$, corresponding to an eigenstate of $s^x$, as expected due to the perfect alignment towards the field deep in the superradiant phase. Importantly, Eq.~\eqref{eqSM:DxMFT} highlights the strong classical-like coherent nature of the photon field in the superradiant regime, as $\Delta_x \propto \langle s^x \rangle$. Such a proportionality persists in the non-Gaussian manifold: while $\Delta_x$ is effectively suppressed by $\lambda$, it still dominates as $\langle s^x \rangle$ grows large (see Fig.~2 of the main text). 

\subsection{Adiabatic impact of ferromagnetic interactions}

As mentioned in the main text, our numerical simulations show that ferromagnetic Ising interactions only adiabatically shift the phase boundary of the standard Dicke model, without altering the nature of the phase on either side.  In other words, the ferromagnetic-normal phase coincides with the normal phase of the standard Dicke Hamiltonian, since the interaction reinforces spin alignment. Here, we provide an analytical explanation for this result. 

In the normal phase, where spins are fully polarized with $\langle \sigma^z \rangle = -1$, correlations are well captured by the mean-field factorization~\cite{dalton2022mag}:
\begin{equation}
    \sigma_i^z \sigma_j^z \approx \sigma_i^z \langle \sigma_j^z \rangle + \langle \sigma_i^z \rangle \sigma_j^z - \langle \sigma_i^z \rangle \langle \sigma_j^z \rangle
    \approx -2 \sigma_i^z - 1\,.
\end{equation}
Thus, the Ising interaction term reduces to an effective Zeeman shift (up to a constant):
\begin{equation}
    \frac{\varepsilon}{2} \sum_i \sigma_i^z - J \sum_{\langle i,j \rangle} \sigma_i^z \sigma_j^z
    \approx \left( \frac{\varepsilon}{2} + 2 J \right) \sum_i \sigma_i^z 
    = \frac{\varepsilon_{\text{eff}}}{2} \sum_i \sigma_i^z ,
\end{equation}
where the renormalized atomic energy splitting is
\begin{equation}
    \varepsilon_{\text{eff}} = \varepsilon + 4J\,.
\end{equation}
This renormalization simply shifts the critical threshold to $g_c(J)=\sqrt{\omega\varepsilon_{\rm eff}}/2$ in agreement with our numerical observations. 

\section{Many-Body effects beyond mean-field theory}

In the standard Dicke model and in the Dicke–Ising model with ferromagnetic interactions, the ground state is well captured by MFT. In this case, the wavefunction factorizes into a direct product of separable Gaussian states that obey Wick’s theorem. The matter and light sectors are strictly decoupled, as shown in Eq.~\eqref{eqSM:pure_variational}: the photon field forms a coherent state, while the spin sector factorizes across sites.

By contrast, when matter–matter interactions are present, as in the Dicke–XXZ model or the Dicke–Ising model with AFM couplings, the Hamiltonian is neither quadratic nor adiabatically connected to one. The ground state must therefore be sought in a broader variational, many-body wavefunction manifold that goes beyond MFT, where Wick’s theorem no longer applies. Such systems develop  genuine spin-spin correlations and entangled spin-photon states. As mentioned in the main text, we employ a variational non-Gaussian class of wavefunctions of the form:
\begin{equation}\label{eqSM:NGS}
    \ket{\Psi_{\rm NGS}} = U_{\rm NGS} 
    \ket{\psi_{\rm ph}} \otimes \ket{\phi}
    \,,
\end{equation}
where $\ket{\phi}$ is a full many-body spin state determined by DMRG with no approximations, and the $U_{\rm NGS}$ is an entangling transformation that correlates photons and spins. In the present work we employ the leading-order form, $U_{\rm NGS}^{(1)}=e^{-\mathcal{S}}$ with $\mathcal{S}=\lambda s^x (a^\dagger - a)$, chosen because it directly reduces the Dicke-type spin-photon coupling in the transformed Hamiltonian. The ground state is variationally obtained by minimizing the average energy described by the wavefunction set in Eq.~\eqref{eqSM:NGS}, referred to as the NGS-DMRG method for short.

\begin{figure}[!t]
    \centering
    \includegraphics[width=0.9\columnwidth]{FigS_DickeIsingEnergy.jpg}\vspace{-3mm}
    \caption{Comparison of energy errors for the Dicke–Ising model at $g=0.5$, obtained from NGS-DMRG (orange) and MFT (gray), relative to exact DMRG results. Calculations are performed on a $N=50$ cluster, with the benchmark DMRG truncating the photon occupation at $M_{\rm max}=40$. The colored bar above the plot denotes the phases.}
    \label{figSM:DickeIsingEnergyComparison}
\end{figure}

This ansatz incorporates nontrivial spin correlations and spin-photon entanglement beyond mean-field theory. In principle, systematic improvement is possible by including higher-order polynomial terms in $\mathcal{S}$, allowing the ansatz to approximate any spin-photon state. However, the variational parameter space grows exponentially with order, rendering such approaches numerically intractable for the system sizes considered. Restricting to the leading-order form thus provides an optimal compromise: it preserves computational tractability while accurately capturing the essential many-body correlations absent from MFT descriptions, as discussed below.

\begin{figure}[!t]
    \centering
    \includegraphics[width=0.9\columnwidth]{FigS_DickeIsingObservables.jpg}\vspace{-3mm}
    \caption{Comparison of (a) average photon occupation and (b) absolute magnetization, obtained from NGS-DMRG results (circles), MFT (gray lines), and exact DMRG results (squares), for the Dicke--Ising model at $g=0.5$ as a function of $J$. Calculations are performed on a $N=50$ cluster, with the benchmark DMRG truncating the photon occupation at $M_{\rm max}=40$. }
    \label{figSM:DickeIsingObservableComparison}
\end{figure}

To highlight the many-body effects captured by the NGS-DMRG method, we benchmark its results against two baselines. First, we compare with MFT, obtained by removing $U_{\rm NGS}$ from Eq.~\eqref{eqSM:NGS} (i.e., setting $\lambda=0$) and restricting the spin state to a direct product. Second, to establish the accuracy of the ansatz, we compare with brute-force DMRG calculations of the full Hamiltonian, performed with a fully converged photon number cutoff $M_{\rm max}$. Because brute-force DMRG becomes rapidly intractable as the photonic Hilbert space grows, we restrict this benchmark to a small cluster, where direct comparison is still feasible.

For the Dicke–Ising model, Fig.~\ref{figSM:DickeIsingEnergyComparison} shows that MFT reproduces the energy reasonably well in the normal phase but deviates strongly in the superradiant regime, where spin–photon entanglement plays a central role. In contrast, the NGS-DMRG results are in excellent agreement with exact DMRG across the full parameter range, with errors on the order of $10^{-5}$. This demonstrates both the accuracy of the variational ansatz and its clear improvement over MFT. Remarkably, these many-body effects do not grow with system size for the Dicke–Ising model: the variational parameters $\xi$ (encoding photon fluctuations) and $\lambda$ (encoding spin–photon entanglement) vanish as $N\to\infty$, indicating that the entanglement between photon and spin sectors becomes asymptotically negligible and the photon field becomes classical in the thermodynamic limit, whereas the spin ground state does not always reduce to a product state.

\begin{figure}[!b]
    \centering
    \includegraphics[width=\columnwidth]{FigS_DickeXXZEnergy.jpg}\vspace{-3mm}
    \caption{Comparison of energy errors for the Dicke–XXZ model at $g=0.1$, obtained from NGS-DMRG (orange) and MFT (gray), relative to exact DMRG results. Calculations are performed on a $N=12$ cluster, with the benchmark DMRG truncating the photon occupation at $M_{\rm max}=15$.}
    \label{figSM:DickeXXZEnergyComparison}
\end{figure}

To reveal how these wavefunction differences manifest in physical observables, we examine the per-site averaged photon occupation and magnetization. As shown in Fig.~\ref{figSM:DickeIsingObservableComparison}, MFT reproduces the trends of the normal phase but fails qualitatively in the superradiant regime: it overestimates the photon number and predicts a smooth crossover to zero, missing the first-order transition, with analogous errors appearing in the magnetization. By contrast, NGS-DMRG quantitatively reproduces the exact results for both observables, demonstrating that the ansatz in Eq.~\eqref{eqSM:NGS} provides a sufficiently accurate description of the spin-photon correlated ground state.

\begin{figure}[!t]
    \centering
    \includegraphics[width=0.9\columnwidth]{FigS_DickeXXZObservables.jpg}\vspace{-3mm}
    \caption{Comparison of (a) average photon occupation and (b) absolute magnetization, obtained from NGS-DMRG results (circles), MFT (gray lines), and exact DMRG results (squares), for the Dicke--XXZ model at $g=0.1$ as a function of $J_z$. Calculations are performed on a $N=12$ cluster, with the benchmark DMRG truncating the photon occupation at $M_{\rm max}=15$.}
    \label{figSM:DickeXXZObservableComparison}
\end{figure}

Many-body correlations play a more pronounced role in the Dicke--XXZ model. In Fig.~\ref{figSM:DickeXXZEnergyComparison}, we compare the ground-state energies obtained from NGS-DMRG and MFT with exact brute-force DMRG results for $N=12$. Unlike the Dicke--Ising case, where MFT was reasonably accurate in the normal phase, here its errors are substantial throughout almost the entire parameter range, exceeding those in Fig.~\ref{figSM:DickeIsingEnergyComparison} by more than an order of magnitude. The NGS-DMRG results, by contrast, remain in excellent quantitative agreement with the exact benchmark, emphasizing that many-body methods are indispensable for this model.

The consequences of these many-body effects are even clearer in observables. As shown in Fig.~\ref{figSM:DickeXXZObservableComparison}(a), MFT strongly overestimates the photon occupation for $J_z<0$ and misidentifies the nature of the phase transition: at $J_z=-1.4$, it predicts a sharp first-order transition, while the exact solution reveals a continuous second-order transition. This mischaracterization is also evident in the magnetization. Such failures are to be expected: the Dicke–XXZ model hosts an intermediate coexistence phase where superradiance scales sublinearly with system size and coexists with XY spin order. This phase is intrinsically a highly correlated quantum state that cannot be represented within the mean-field manifold. Because the in-plane symmetry prevents the suppression of quantum fluctuations, they remain robust even in the strong light–matter coupling regime, ensuring that MFT cannot become asymptotically exact.
In contrast, NGS-DMRG closely tracks the exact results across all observables, with only a minor underestimation of the photon number ($\sim10^{-4}$). This confirms that the variational NGS-DMRG ansatz remains accurate and reliable for describing the highly correlated Dicke--XXZ ground state.